\documentclass[useAMS,usenatbib,usegraphicx]{mn2e}

\title[The disc-jet coupling in Aql X-1]{The disc-jet coupling in the neutron star X-ray binary Aquila X-1}
\author[V. Tudose et al.]{V. Tudose,$^{1,2,3}$\thanks{E-mail:
tudose@astron.nl (VT)}\thanks{Present address: Netherlands Institute
for Radio Astronomy, Oude Hoogeveensedijk 4, 7991 PD Dwingeloo, the Netherlands} R.P. Fender,$^{4,1}$ M. Linares,$^{1}$ 
D. Maitra$^{1}$ and M. van der Klis$^{1}$  \\
$^{1}$Astronomical Institute `Anton Pannekoek', University of Amsterdam, Kruislaan 403, 1098 SJ Amsterdam, the Netherlands\\
$^{2}$Astronomical Institute of the Romanian Academy, Cutitul de Argint 5, RO-040557 Bucharest, Romania\\
$^{3}$Research Center for Atomic Physics and Astrophysics, Atomistilor 405, RO-077125 Bucharest, Romania\\
$^{4}$School of Physics and Astronomy, University of Southampton,
Highfield, Southampton SO17 1BJ}

\begin{document}

\date{Accepted 2009 August 24. Received 2009 August 24; in original form 2008 November 26}

\pagerange{\pageref{firstpage}--\pageref{lastpage}} \pubyear{2009}

\maketitle

\label{firstpage}

\begin{abstract}

We study the accretion/ejection processes (i.e. disc/jet coupling) in the neutron star X-ray binary Aquila X-1 via a multi-wavelength approach. 
We use in the radio band the publicly available Very Large Array archive containing observations of the object between 1986--2005, in the X-ray 
band the archival Rossi X-ray Timing Explorer data (Proportional Counter Array and High Energy X-ray Timing Experiment) between 1997--2008, and 
in optical (R band) observations with the Small and Moderate Aperture Research Telescope System recorded between 1998--2007. In the combined data set 
we find three outbursts for which quasi-simultaneous radio, optical (R band) and X-ray data exist and focus on them to some extent. We provide 
evidence that the disc/jet coupling in Aquila X-1 is similar to what
has been observed in black hole X-ray binaries, at least from the point of 
view of the behaviour in the hardness-intensity diagrams (the hysteresis effect included), when the phenomenology of the jet is taken into account. 
Although based on a very small number of observations, a radio/X-ray correlation seems to exist for this system, with a slope of 
$\alpha$=0.40 $\pm$ 0.07 ($F_{radio} \propto F_{X}^{\alpha}$), which is different than the slope of $\alpha$=1.40 $\pm$ 0.25 found for another 
atoll source, 4U 1728-34, but interestingly enough is relatively close to the values obtained for several black hole X-ray binaries. No significant correlation 
is found between the radio and optical (R band) emissions. We also report a significant drop in the radio flux from Aql X-1 above
an X-ray flux of $\sim$  5 $\times$ 10$^{-9}$ erg cm$^{-2}$
s$^{-1}$. This behaviour, also reported in the neutron star X-ray binary 4U 1728-34, may be analogous to the
suppression of radio emission in black hole X-ray binaries in bright,
soft X-ray states. It suggests that from this point of view neutron star X-ray binaries can mimic the behaviour of black hole 
X-ray binaries in suppressing the jet in soft/disc-dominated X-ray states.

\end{abstract}

\begin{keywords}
accretion, accretion discs -- stars: individual: Aquila X-1 -- ISM: jets and outflows 
-- radiation mechanisms: non-thermal -- X-rays: stars.
\end{keywords}	

\section{Introduction}

The low-mass X-ray binary (LMXRB) Aquila X-1 (Aql X-1) is a recurrent soft X-ray transient which shows quasi-periodic outbursts 
about once a year \citep{Pri84,Kit93,Sim02}. The compact object in the system is a neutron star as implied by the detection of type I 
X-ray bursts (e.g. \citealt{Koy81,Cze87,Yu99}). The companion is a main sequence star likely of the spectral type K7 \citep{Che99}. 
The X-ray timing and spectral properties of 
Aql X-1 place it in the class of atoll sources \citep{Rei00} and recent observations of coherent pulsations in the persistent X-ray 
emission \citep{Cas08} makes it the ninth accretion-powered millisecond
X-ray pulsar (AMP) known to date. The atolls show two main X-ray states \citep{Has89,Kli89}, 
identifiable in the colour-colour diagrams: a softer, ``banana state''
(BS), and a harder, ``island state'' (IS). In order to 
accommodate later observations of harder X-ray spectra a new state was
defined (e.g. \citealt{Pri97}): the ``extreme island state''
(EIS). From the point of view of the X-ray spectral and timing properties, atoll sources (especially in the EIS) 
share many properties with the black hole XRBs (BHXRBs) in the low-hard
state (e.g. \citealt{Kli94,Oli98,Ber98}).

Aql X-1 also exhibits 
X-ray burst oscillations (e.g. \citealt{Zha98}) and kHz quasi-periodic oscillations (QPOs), both lower kHz QPO (e.g. 
\citealt*{Cui98,Men01,Rei04}), and upper kHz QPO (a single detection, \citealt*{Bar08}).

The optical counterpart has in quiescence a V band magnitude of 21.60 and is only 0\farcs48 away from a contaminating star 
of V band magnitude 19.42 \citep{Che99} thus complicating the optical studies.

Besides the orbital period close to 19 h \citep*{Che91,Che98,Sha98,Wel00}, little is known about the other parameters of the system.
The inclination of the orbit is poorly constrained, with values ranging between 36$\degr$ and 70$\degr$ \citep*{Sha97,Gar99,Wel00}.
The distance to Aql X-1 is not well known, but different estimates place it in the range 4--6.5 kpc \citep{Che99,Rut01,Jon04}.

Reports of radio observations of Aql X-1 are extremely scarce in the
literature: \cite*{Hje90,Rup04,Rup05}. To date, only five radio detections
have been documented, all obtained by the VLA during outbursts: four at 8.4 GHz (two in 1990
August and two in 2004 May), and one at 4.9 GHz (in 2005 April).  This is not surprising since 
the atoll sources are quite difficult to observe given their sub-mJy flux density levels at cm wavelengths even during outbursts 
\citep{Fen01b}. Basically, although they represent the bulk of the XRBs population \citep{Fen06}, only a handful of atoll sources have 
been detected in radio (e.g. \citealt{Hje95,Mar98,Mig03,Mig04,Moo00}). In the case of the atoll 
source 4U 1728-34 a correlation was found between the radio and X-ray
fluxes in the hard state \citep{Mig03,Mig06}, resembling the relation 
established for BHXRBs in the low-hard state (\citealt*{Cor03,Gal03,Gal06,Cor08}, but see also \citealt{Xue07}), but with a steeper gradient.

\section{Observations}

\begin{table*}
 \centering
  \caption{The VLA 8.4 GHz radio detections and a selection of upper
limits (3 $\sigma$) together with the corresponding quasi-simultaneous observations in 
optical (R band) and X-ray (PCA/RXTE and HEXTE/RXTE) bands.}
  \begin{tabular}{@{}lccccccccc}
  \hline \hline
 Date & MJD & Flux density & MJD & Flux density & MJD & Flux & MJD &
Count rate & X-ray  \\
      & radio & radio    & optical & optical   & PCA & PCA        & HEXTE & HEXTE & state  \\
& [day]   & [$\mu$Jy] & [day]   & [$\mu$Jy] & [day]   & [10$^{-9}$ erg
cm$^{-2}$ s$^{-1}$] & [day] & [c s$^{-1}$] & \\
\hline
1990 August    & 48111.23 & 415 $\pm$ 38 & - & - & - & - & - & - & -  \\
1990 August    & 48129.23 & 150 $\pm$ 21 & - & - & - & - & - & - & -  \\
2002 March     & 52334.62 & $<$ 69       & 52334.39 & 372 $\pm$ 12 & 52335.04 & 5.57 $\pm$ 0.56 & 52332.92 & 1.1 $\pm$ 0.3  & BS  \\
2002 March     & 52355.67 & 179 $\pm$ 26 & 52354.87 & 302 $\pm$ 10 & 52355.84 & 1.40 $\pm$ 0.20 & 52356.69 & 2.3 $\pm$ 0.4  & IS  \\
2002 April     & 52379.60 & $<$ 68       &    -     &      -       &     -    &       -           &     -    &       -        & -   \\
2002 May       & 52399.58 & $<$ 57       & 52400.82 & 16  $\pm$ 2  &     -    &       -           &     -    &       -        & -   \\
2002 May       & 52413.52 & $<$ 63       & 52414.29 & 1   $\pm$ 2  &     -    &       -           &     -    &       -        & -   \\
2004 March     & 53073.71 & $<$ 183      &    -     &       -      & 53074.90 & 1.26 $\pm$ 0.11 & 53074.91 & 11.3 $\pm$ 0.3 & EIS \\
2004 March     & 53077.48 & $<$ 147      &    -     &       -      & 53078.84 & 1.06 $\pm$ 0.11 & 53078.84 & 8.7  $\pm$ 0.3 & EIS \\
2004 March     & 53085.56 & $<$ 174      &    -     &       -      & 53085.93 & 1.06 $\pm$ 0.11 & 53085.93 & 9.2  $\pm$ 0.3 & EIS \\
2004 March     & 53091.56 & $<$ 150      & 53092.38 & 297 $\pm$ 10 & 53088.90 & 0.84 $\pm$ 0.21 &     -    &       -        & EIS \\ 
2004 May       & 53144.50 & 179 $\pm$ 22 & 53144.36 & 610 $\pm$ 19 & 53144.63 & 1.44 $\pm$ 0.14 & 53143.65 & 16.1 $\pm$ 0.7 & EIS \\
2004 May       & 53151.42 & 216 $\pm$ 19 & 53151.35 & 793 $\pm$ 24 & 53151.71 & 1.97 $\pm$ 0.13 & 53151.71 & 17.1 $\pm$ 0.4 & EIS \\
2004 June      & 53162.37 & 205 $\pm$ 19 & 53162.38 & 540 $\pm$ 17 & 53162.67 & 2.41 $\pm$ 0.20 & 53162.67 & 17.3 $\pm$ 0.4 & EIS \\
2004 June      & 53170.35 & 274 $\pm$ 22 & 53169.40 & 440 $\pm$ 14 & 53170.40 & 3.89 $\pm$ 0.51 & 53172.37 & 0.6  $\pm$ 0.5 & BS  \\
2004 June      & 53176.29 & $<$ 114      & 53177.15 & 121 $\pm$ 5  & 53176.92 & 0.40 $\pm$ 0.04 & 53176.92 & 4.2  $\pm$ 0.4 & EIS \\
2004 June      & 53183.29 & $<$ 126      & 53182.27 & 51  $\pm$ 3  & 53182.75 &    $<$ 0.04  &     -    &       -        & EIS \\
2004 July      & 53187.23 & $<$ 117      & 53187.30 & 15  $\pm$ 2  &     -    &       -           &     -    &       -        & -   \\ 
\hline \hline
\end{tabular}
\end{table*}

\begin{table*}
 \centering
  \caption{The VLA 4.9 GHz radio detections and a selection of upper
limits (3 $\sigma$) together with the corresponding quasi-simultaneous observations in 
optical (R band) and X-ray (PCA/RXTE and HEXTE/RXTE) bands.}
  \begin{tabular}{@{}lcccccccccc}
  \hline \hline
 Date & MJD & Flux density & MJD & Flux density & MJD & Flux & MJD &
Count rate & X-ray  \\
      & radio & radio    & optical & optical   & PCA & PCA        & HEXTE & HEXTE & state  \\
& [day]   & [$\mu$Jy] & [day]   & [$\mu$Jy] & [day]   & [10$^{-9}$ erg
cm$^{-2}$ s$^{-1}$] & [day] & [c s$^{-1}$] & \\
\hline
1990 August    & 48111.23 & 289 $\pm$ 24 & - & - & - & - & - & - & - \\
2002 March     & 52355.67 & 154 $\pm$ 17 & 52354.87 & 302 $\pm$ 10 & 52355.84 & 1.40 $\pm$ 0.20 & 52356.69 & 2.3   $\pm$ 0.4 & IS  \\
2002 April     & 52379.58 & $<$ 105      &   -      &     -        &     -    &         -         &    -     &        -        & -   \\
2002 May       & 52399.54 & $<$ 93       & 52400.82 & 16  $\pm$ 2  &     -    &         -         &    -     &        -        & -   \\
2004 May       & 53151.42 & $<$ 255      & 53151.35 & 793 $\pm$ 24 & 53151.71 & 1.97 $\pm$ 0.13 & 53151.71 & 17.1  $\pm$ 0.4 & EIS \\
2004 June      & 53162.37 & $<$ 276      & 53162.38 & 540 $\pm$ 17 & 53162.67 & 2.41 $\pm$ 0.20 & 53162.67 & 17.3  $\pm$ 0.4 & EIS \\
2004 June      & 53170.35 & $<$ 246      & 53169.40 & 440 $\pm$ 14 & 53170.40 & 3.89 $\pm$ 0.51 & 53172.37 & 0.6   $\pm$ 0.5 & BS  \\
2004 June      & 53176.35 & $<$ 174      & 53177.15 & 121 $\pm$ 5  & 53176.92 & 0.40 $\pm$ 0.04 & 53176.92 & 4.2   $\pm$ 0.4 & EIS \\
2004 June      & 53183.29 & $<$ 282      & 53182.27 & 51  $\pm$ 3  & 53182.75 &     $<$ 0.04 &    -     &        -        & EIS \\
2005 March     & 53446.71 & $<$ 159      &   -      &     -        &     -    &         -         &    -     &        -        & -   \\  
2005 April     & 53465.54 & 245 $\pm$ 29 & 53465.34 & 444 $\pm$ 14 & 53465.66 & 1.13 $\pm$ 0.11 & 53465.66 & 10.4  $\pm$ 0.4 & EIS \\
2005 April     & 53472.50 & 317 $\pm$ 30 &   -      &     -        & 53472.81 & 2.70 $\pm$ 0.47 & 53472.81 & 19.5  $\pm$ 0.4 & EIS \\
2005 May       & 53494.52 & $<$ 156      & 53492.33 & 228 $\pm$ 8  & 53493.79 & 0.84 $\pm$ 0.07 & 53493.79 & 6.7   $\pm$ 0.3 & EIS \\
2005 May       & 53506.48 & $<$ 207      & 53508.31 & 107 $\pm$ 5  & 53505.85 & 0.09 $\pm$ 0.01 &    -     &        -        & EIS \\
\hline \hline
\end{tabular}
\end{table*}

\subsection{Radio}
We have analyzed all the public Very Large Array (VLA) radio data between 1986 February and 2005 December 
containing observations of Aql X-1 (close to 100 epochs). The majority of the 
observations were carried out at multiple frequencies, particularly at 4.9 and 8.4 GHz. Other frequencies, namely 1.5 and 15.0 GHz, 
were observed much less frequently. The bandwidth used was 100 MHz. All the runs were performed after reports of increased activity from Aql X-1 
mainly at optical and X-ray wavelengths, and thus generally trace the outburst history of the object. Given the target of opportunity 
nature of the observations, the array configurations were varying between runs, and the effective 
observing time ranged roughly between 5 and 60 min, with the median around 10--20 min. Primary calibrators were either 3C286 or 3C48. 
As secondary calibrators we used, upon availability, J1950+081 or J1925+211 for the vast majority of the epochs, J1922+155, J1939-100, 
J2007+404, J1851+005, or J1820-254. The data were calibrated in AIPS04 using standard procedures as highlighted in \cite{Dia95}. The imaging was done in DIFMAP
2.4e \citep{She97} by progressively cleaning the residual map using the
CLEAN algorithm of \cite{Hog74}. The {\it uv-}plane fitting was carried out
with elliptical Gaussians within the same software. 

We detected Aql X-1 at more than 3$\sigma$ levels on 7 epochs at 8.4 GHz (Table 1) and 4 epochs at 4.9 GHz (Table 2). In two occasions 
the detections were 
quasi-simultaneous at both frequencies. At least in the cases of some outbursts it is likely that the lack of detection in the radio band is 
partly due to the very short integration times which generated relatively high noise levels. In Tables 1 and 2 we also include the 
upper limits (3 $\sigma$) of the flux densities for the observations pertaining to outbursts for which there are radio detections.

When detected, the object was unresolved. All the flux densities 
reported here are determined via {\it uv-}plane fitting and the corresponding errors are estimated based on the rms noise of the images 
and the differences between flux densities measured in the {\it uv-}plane using different initial conditions. Two contaminating sources 
are present in the field of view, $\sim$40 and $\sim$290 arcsec away from the position of Aql X-1, and affect the quality of the 
images in compact configurations at lower frequencies (i.e. 4.9 GHz). However, this should not significantly influence 
the flux measurements. Our values are in agreement, within the errors,
with those reported by \cite{Hje90} and \cite{Rup04,Rup05}. The only exception is 
the data at 4.9 GHz from 2005 April (MJD 53465.54) when we clearly detect the target while in the previous work only a marginal detection 
is noted. This inconsistency comes probably from a misidentification of
the source on this particular epoch. Generally speaking, the radio flux density levels of Aql
X-1 are relatively low. At times, noise features are present in the
radio maps and can be misidentified as the target. This could have happened
on this instance we mention.

\subsection{X-ray}

We used archival Rossi X-ray Timing Explorer (RXTE) Proportional Counter Array (PCA) and High Energy X-ray Timing Experiment (HEXTE) 
data of Aql X-1 available in NASA's High Energy Astrophysics Science Archive Research Center (HEASARC). 

We filtered out data taken when the spacecraft was pointing away from
the source (offset $>$ 0.02 degrees), too close to the Earth (elevation
$<$ 10 degrees) or during and up to 15 minutes after the South Atlantic
Anomaly passages. All
RXTE products were extracted using the standard HEASOFT tools (version 6.5).

In the case of the PCA, the count rates were extracted 
using the Standard 2 data of all active PCUs (with 16 s time resolution) and averaged over
each observation. Following previous work (e.g. \citealt{Rei04}) we
used the following energy bands: A: 2.0--3.5 keV; B: 3.5--6.0 keV; C: 6.0--9.7 keV; D: 9.7--16.0 keV. The colours and 
intensity were defined as: soft colour SC=B/A, hard
colour HC=D/C, intensity I=A+B+C+D. We applied dead-time corrections and
subtracted the background using {\it pcabackest} (version 3.6),
the latest models and updated SAA history
files\footnote{http://heasarc.gsfc.nasa.gov/docs/xte/pca\_news.html}.
Following the recommended procedure\footnote{http://heasarc.gsfc.nasa.gov/docs/xte/pca\_news.html\#quick\_table}, we used the faint source model
when the source was below $\sim$40 c/s/PCU and the bright source model otherwise.

The PCA gains changed several times during the lifetime of RXTE,
generating different ``gain epochs''. On top of these sudden changes, a
long term gain decay is also present \citep{Jah06}. In order to confidently compare
count rates measured at different epochs, months to years apart, we
normalized the count rates to those of the Crab nebula, which is known
to be a steady X-ray source. The conversion from count rate to Crab was done using the value of the
Crab rate closest in time and within the same PCA gain epoch (e.g. \citealt{Kuu94,Str03}). To obtain
the unabsorbed flux in the 2--10 keV band we extracted background and
dead-time corrected PCA spectra using only PCU 2, as it has been the most active and best calibrated throughout the RXTE mission.
We estimated the background spectra and created response matrices using
{\it pcabackest} (as explained above) and {\it pcarsp}, respectively
(both part of the same HEASOFT v. 6.5 package), and added a 1\%
systematic error to all channels and grouped them in order to have a minimum of 20 counts per energy bin.
We then fitted the spectra within XSPEC 11.3.2ag
using the following model: wabs*(bbody+diskbb+powerlaw), and fixing the
hydrogen column density to 0.5 $\times$ $10^{22}$ cm$^{-2}$
\citep{Chu01}. In one case the system was in quiescence and we
determined the corresponding upper limit on the flux using an absorbed power-law to fit
the unmodeled background (diffuse emission).

In addition, we extracted 20--200 keV dead-time and background
corrected count rates from the HEXTE standard mode data, and averaged
them within each observation. We averaged both rocking directions in
order to measure the background spectrum during each observation. For
the analysis in section 6 we used cluster B data (excluding detector 2,
which was damaged in March 1996), while for the rest of the paper
cluster A data (all detectors). This choice was determined by the
periods of anomalous rocking of cluster A, which otherwise might have affected the long-term analysis done in section 6.
The overall gain and response matrix of the HEXTE detectors are
independent of time and hence no Crab nebula normalization was applied
to the HEXTE data.

\subsection{Optical}

For the current work we used part of the data from the long-term
monitoring campaign on Aql X-1 reported in \cite{Mai08}. Details of the
data reduction and analysis can be found in section 2.1 of the
aforementioned reference. 

The R-band observations were 
taken using the ANDICAM\footnote{http://www.astro.yale.edu/smarts/ANDICAM} instrument on the 1.0 m and 1.3 m Small and Moderate Aperture 
Research Telescope System (SMARTS) at Cerro Tololo Inter-American Observatory (CTIO). The data reduction was done via the standard IRAF 
pipeline developed for ANDICAM. 

Around 83 percent of the quiescent R-band flux is estimated to come from the contaminating star \citep{Wel00} and was thus subtracted 
from the total observed flux from all the observations.

For magnitude to flux conversion it was assumed that R=0 magnitude corresponds to 3064 Jy \citep*{Cam85,Bes98}. In \cite{Che99} the colour 
excess of Aql X-1 was estimated as E(B-V)=0.5 $\pm$ 0.1. The R-band extinction $A_R$ was determined using the standard total to selective 
extinction coefficient $A_V/{E(B-V)}$=3.1 and the interstellar extinction law $A_R/A_V$=0.748 \citep{Rie85,Car89}.

\section{Multi-wavelength light curves}

For the three radio detections in 1990 August no multi-wavelength information was available and therefore we do not discuss 
these data here. We only note that during the quasi-simultaneous detections at 4.9 and 8.4 GHz on MJD 48111.23 the spectrum between these 
frequencies was inverted (i.e. $\alpha \geq 0$ where $F_{\nu} \propto \nu^{\alpha}$). 

\begin{center}
\begin{figure}
  \includegraphics[scale=0.43]{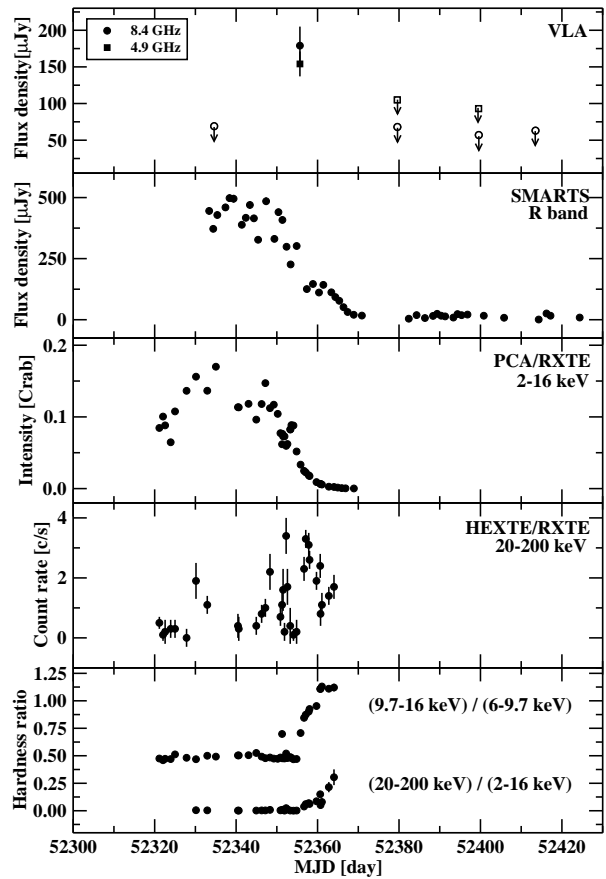}
  \caption{Multi-wavelength light curves for the outburst of Aql X-1 from 2002 March and the variation of the hardness ratio between two PCA bands 
(9.7--16 keV and 6--9.7 keV), and the HEXTE (20--200 keV) and PCA (2--16 keV) bands. If not appearing in the plots, the errors are smaller than 
the size of the points. The open symbols in the radio light curve
denote 3 $\sigma$ upper limits.}
\end{figure}
\end{center}

\begin{center}
\begin{figure}
  \includegraphics[scale=0.43]{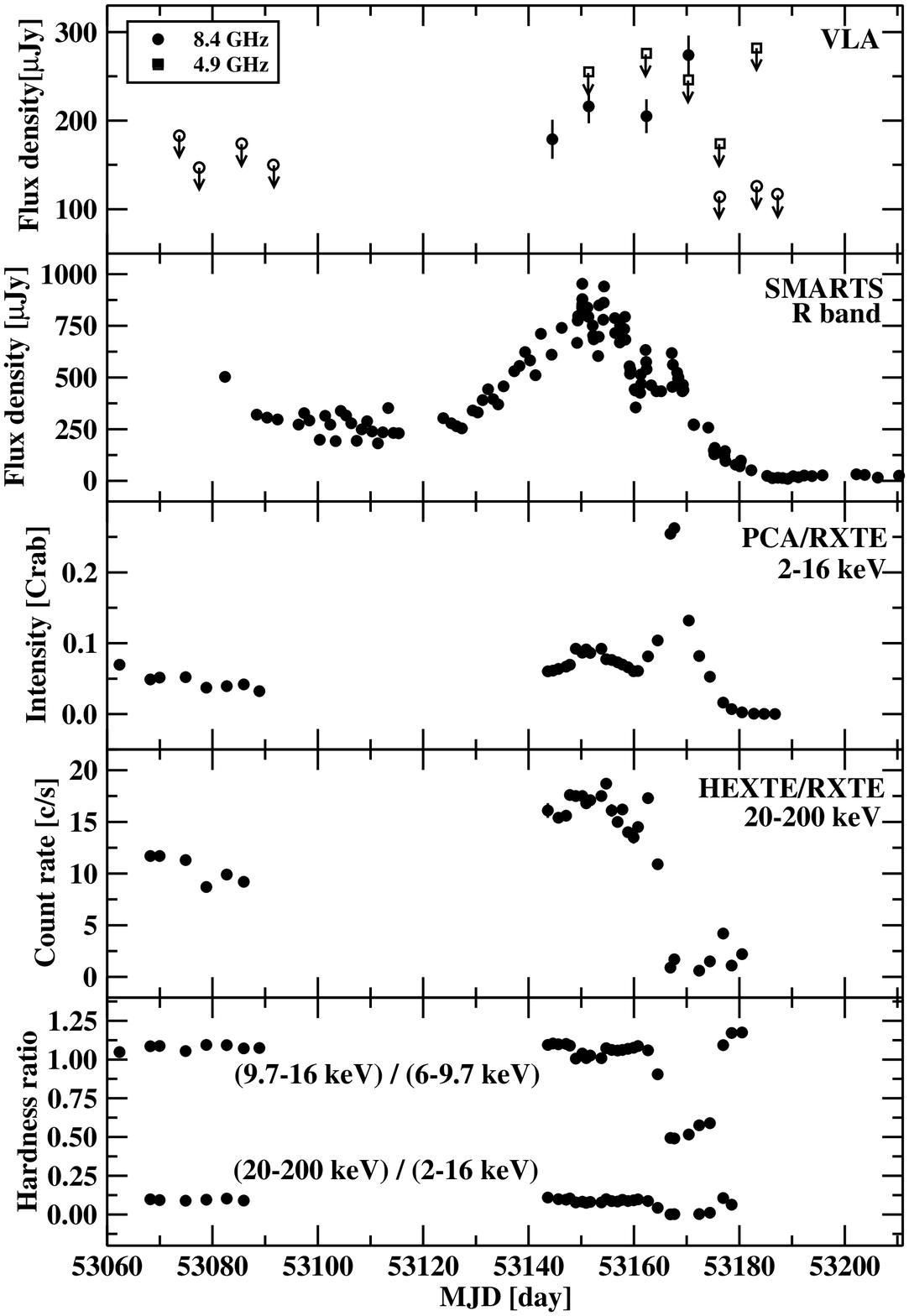}
  \caption{Multi-wavelength light curves for the outburst of Aql X-1 from 2004 May--June and the variation of the hardness ratio between two PCA bands 
(9.7--16 keV and 6--9.7 keV), and the HEXTE (20--200 keV) and PCA (2--16 keV) bands. If not appearing in the plots, the errors are smaller than 
the size of the points. The open symbols in the radio light curve
denote 3 $\sigma$ upper limits.}
\end{figure}
\end{center}

\begin{center}
\begin{figure}
  \includegraphics[scale=0.43]{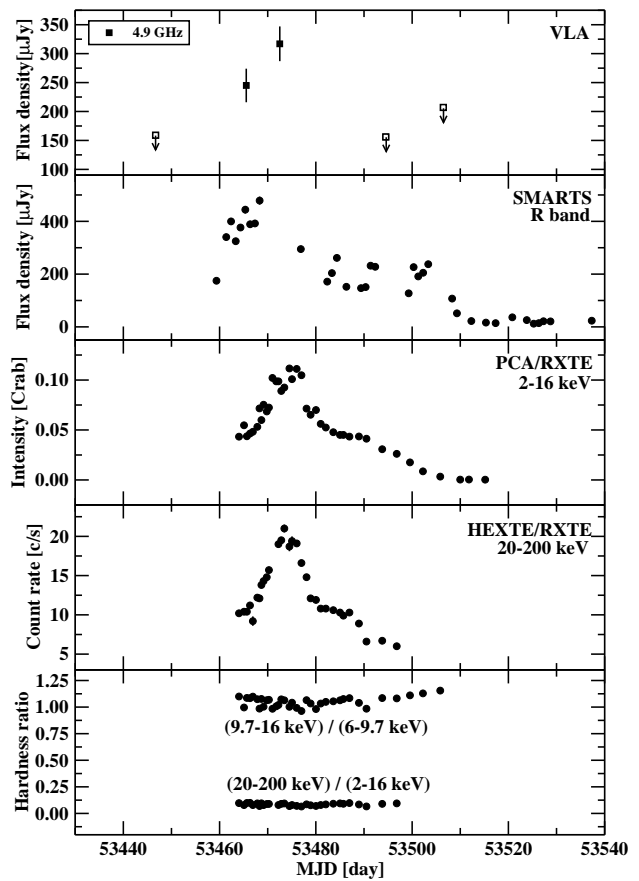}
  \caption{Multi-wavelength light curves for the outburst of Aql X-1 from 2005 April and the variation of the hardness ratio between two PCA bands 
(9.7--16 keV and 6--9.7 keV), and the HEXTE (20--200 keV) and PCA (2--16 keV) bands. If not appearing in the plots, the errors are smaller than 
the size of the points. The open symbols in the radio light curve
denote 3 $\sigma$ upper limits.}
\end{figure}
\end{center}

\subsection{2002 March outburst}

During this outburst (Fig. 1) Aql X-1 was detected in radio quasi-simultaneously at 4.9 and 8.4 GHz and showed then an inverted spectrum. The optical 
and PCA data clearly indicate that the source was passing through an active state. It is also apparent that the beginning of the outburst 
was missed. The level of the HEXTE count rate is low during the $\sim$40 days of monitoring and tends to rise slightly towards the end 
of the data set. As indicated by the hardness ratio, most of the period the system was in a soft state which it left suddenly and transited 
quickly to a hard state around MJD 52355. The radio detections are coincident with this X-ray state transition. 

\subsection{2004 May--June outburst}

This is so far, relatively speaking, the most well covered outburst of Aql X-1 in the radio band (Fig. 2). The source was detected on 4 occasions and 
only at 8.4 GHz, however the lack of detection at 4.9 GHz might be partly attributable to the high rms noise due to the short integration times. 
On the epoch with the highest 8.6 GHz flux density detected, the
quasi-simultaneous upper limit at 4.9 GHz allows us to constrain the spectrum in this 
frequency range to an inverted one. The optical light curve is well sampled and the PCA data reveal a major outburst preceded by a smaller 
flaring event at about MJD 53150. After an apparent peak, the HEXTE registered a drop in the count rate soon after the onset 
of the intensity increase 
in the PCA band. This has been observed before in Aql X-1 during two previous outbursts, by \cite{Yu03}. The hardness ratio diagram indicates that 
during the active period the system made a series of 
X-ray state transitions, from hard to soft and back to hard. A similar
behaviour is observed for the smaller flare, when the hardness ratio
decreased by 10 percent, with the difference 
that in this case the system never reached the soft state, but only temporarily softened its spectrum. The increases in the radio flux density 
seem to trace the intensity enhancements in the PCA band. 

\subsection{2005 April outburst}

For this outburst (Fig. 3), observations at 8.4 GHz were not available. Aql X-1 was detected twice at 4.9 GHz. The optical data is quite sparse and 
the onset of the burst was not covered well. The PCA and HEXTE intensity peaks are more or less coincident. The hardness ratio diagram 
reveals that the system never left the hard X-ray state during the active period. 

\section{Hardness--Intensity diagrams}

\begin{center}
\begin{figure*}
  \includegraphics*[scale=0.3]{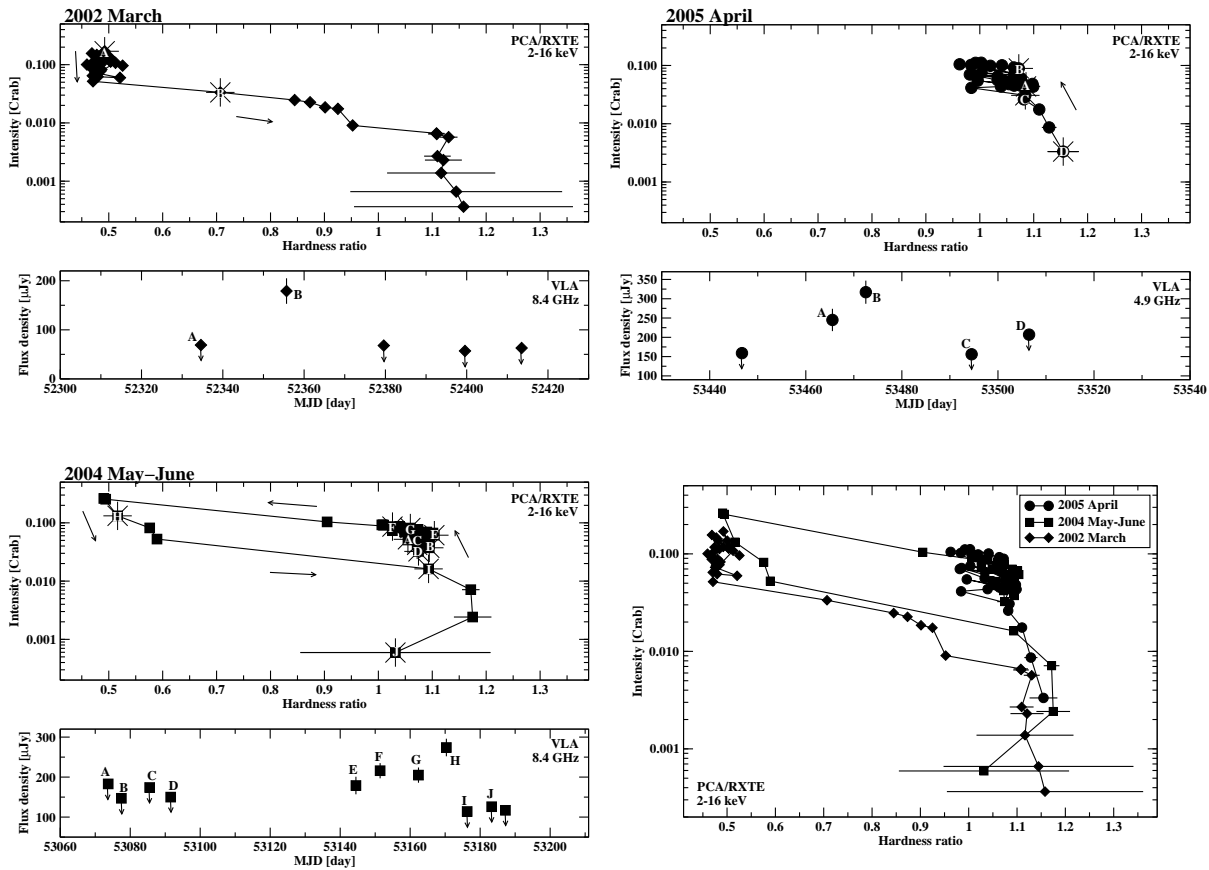}
  \caption{HIDs for the outbursts of Aql X-1 from 2002 March (upper-left; X-ray data between MJD 52321.1--MJD 52366.9), 2004 May--June (bottom-left; 
X-ray data between MJD 53053.8--MJD 53182.8) and 2005 April (upper-right; X-ray data between MJD 53464.0--MJD 53505.8). The cumulative 
HID for the all three outbursts is represented in the bottom-right panel. The
hardness is the ratio of the count rates in the PCA bands 9.7--16.0 keV and 
6.0--9.7 keV. The points highlighted in each HID are shown in the corresponding radio light curve. The other points do not correspond to quasi-simultaneous 
radio and X-ray observations.}
\end{figure*}
\end{center}

\begin{center}
\begin{figure}
  \includegraphics[scale=0.33,angle=-90]{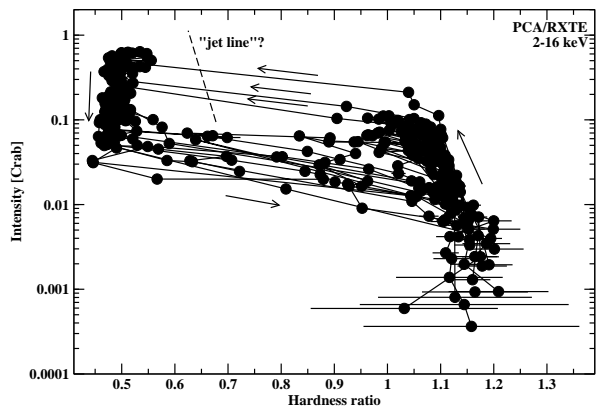}
  \caption{The cumulative HID for all the outbursts of Aql X-1 observed with the PCA/RXTE between 1997--2008. A hysteresis phenomenon is clearly 
at work in Aql X-1. The hardness is the ratio of the count rates in the PCA bands 9.7--16.0 keV and 6.0--9.7 keV. Superimposed (see section 7 for details) 
is the ``jet line'' of e.g. \citet{Fen04}.}
\end{figure}
\end{center}

The hardness-intensity diagrams (HIDs) corresponding to the three outbursts are represented in Fig. 4. The hardness ratio is the hard colour defined 
in section 2.2. Note that the scales for all the HIDs are identical. The system is tracing the HIDs counter-clockwise.

In 2002 March (Fig. 4, upper-left) the radio upper limit denoted ``A'' corresponds to a soft X-ray state of the system at a relatively high X-ray 
intensity. This situation is reminiscent of the similar behaviour observed in BHXRBs, when the radio emission is ``quenched'' above some X-ray 
flux level. We postpone a further discussion of the issue to section 6. The system stays in a soft X-ray state for a while and then makes 
a fast transition to a hard state. The radio detection (``B'') caught the system during this transitional phase. 

In 2004 May--June (Fig. 4, bottom-left) the system traced a full cycle
in the diagram. It looks similar to the HIDs 
observed in BHXRBs (e.g. \citealt{Fen04}) and recently in a dwarf nova \citep{Kor08}. The first four radio non-detections (``A'' to ``D'') as well 
as the first three detections (``E'' to ``G'') correspond to a hard X-ray state. Then suddenly the system jumped to a soft X-ray state and 
the radio flux density increased significantly (``H''), by more than 30
percent (with the errors on the individual measurements of less than 10
percent). On a timescale comparable to the rise of the radio flux
density, the system dropped bellow 
the radio detection limit, in a hard X-ray state (``I'', ``J''). 

In 2005 April (Fig. 4, upper-right) the system stayed constantly in a hard X-ray state. It experienced what is sometimes called a hard outburst 
(e.g. \citealt{Mig06}). This has been observed in Aql X-1 on a few occasions (e.g. \citealt{Rod06}).

Fig. 4 clearly illustrates hysteresis in Aql X-1 (the X-ray luminosity at the transition 
from hard to soft state is different, i.e. higher, than the luminosity at the transition from soft to hard state). This kind of behaviour was previously 
noted for Aql X-1 \citep*{Mac03,Mai04,Gla07}. 
A HID with all the PCA data for all the X-ray outbursts of Aql X-1 between 1997 and 2008 clearly shows this (Fig. 5). Three major flares have been relatively 
well sampled and the corresponding hard to soft transitions in Fig. 5
are therefore not observationally biased. 
This phenomenon of hysteresis has been seen in BHXRBs (e.g. \citealt{Rod03,Ros04,Bel05,Dun08}) suggesting a common origin for the state transitions in NSXRBs 
and BHXRBs. In Fig. 5 the hard to soft transitions might appear to be significantly quicker than the soft to hard ones. This is an artifact due to the 
adding of the data together as some minor outbursts also follow tracks in
the region between 0.01--0.10 Crab. Whether or not they trace a scaled down 
version of the large-scale diagram is not clear, but the superposition
creates a region with a higher density of points, which
could falsely be interpreted as symptoms of slower soft to hard transitions. 

\section{Colour--Colour diagrams}

\begin{center}
\begin{figure*}
  \includegraphics*[scale=0.3]{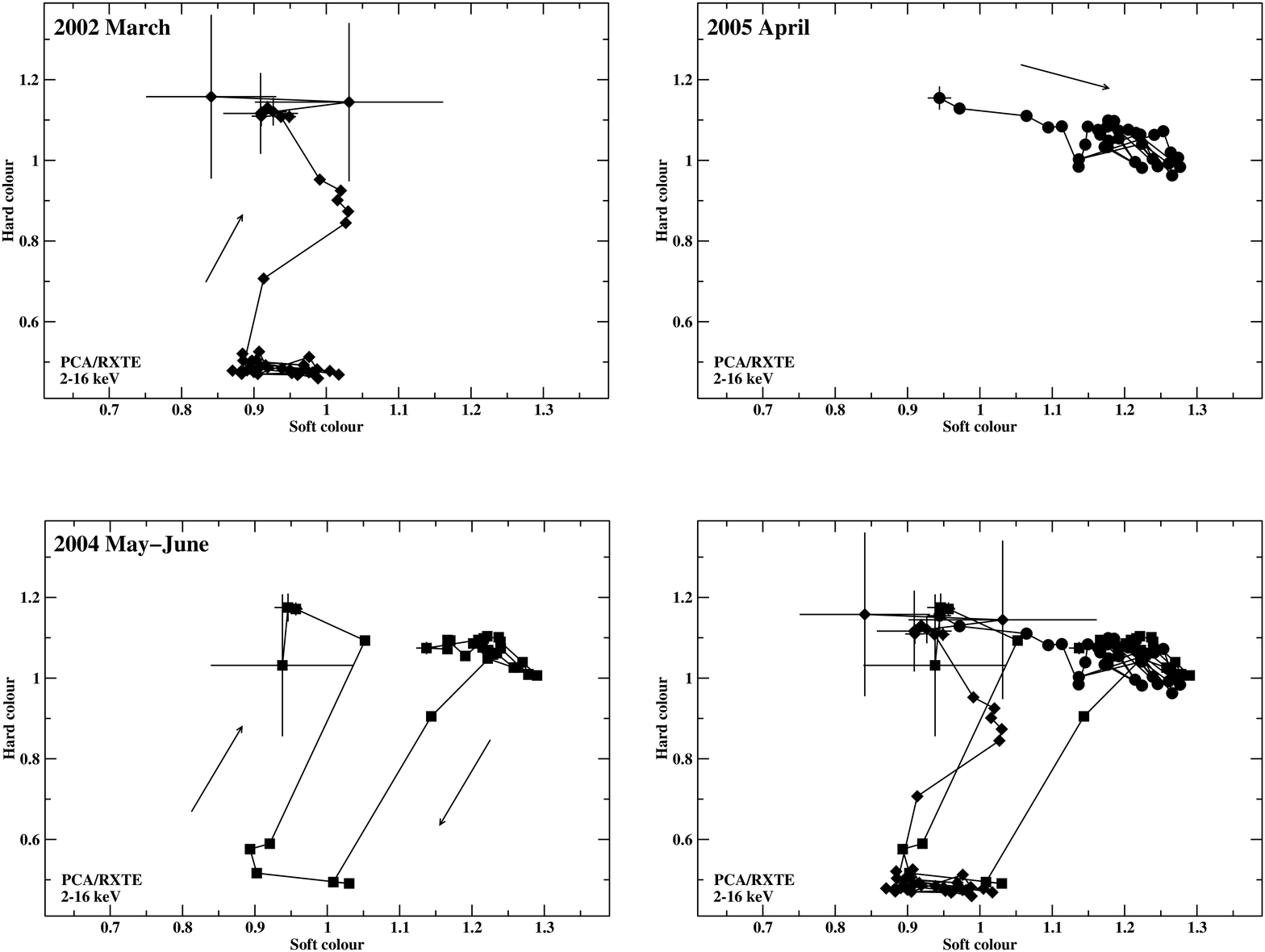}
  \caption{CCDs for the outbursts of Aql X-1 from 2002 March (upper-left; X-ray data between MJD 52321.1--MJD 52366.9), 2004 May--June (bottom-left; 
X-ray data between MJD 53053.8--MJD 53182.8) and 2005 April (upper-right; X-ray data between MJD 53464.0--MJD 53505.8). The cumulative 
CCD for the all three outbursts is represented in the bottom-right. The
soft colour is the ratio between the count rates in the 3.5--6.0 keV and 
2.0--3.5 keV bands and the hard colour is the ratio between the count rates in the 9.7--16.0 keV and 6.0--9.7 keV bands.}
\end{figure*}
\end{center}

The colour-colour diagrams (CCDs) for the three outburst studied here in detail are shown in Fig. 6. The system is tracing the CCDs clockwise. 

Already in Fig. 6 the CCD for the three outbursts hints that Aql X-1 seems to spend most of the time in two main X-ray states during the active periods: 
a harder state roughly with HC$>$0.9, and a softer state with HC$<$0.6 and a narrower interval of variation of SC. This is clearly confirmed by the 
CCD for all the outbursts observed with the PCA between 1997-2008 (Fig. 7). Besides the two main X-ray states, a transitional one is now evident, linking 
the two. This CCD is, not surprisingly, entirely consistent with the CCD built by \cite{Rei04} with all the PCA data between 1997--2002 (note that 
we averaged the data over each observation). Following these authors and the nomenclature for atoll sources, we have identified the harder 
X-ray state as the EIS, the softer one as the BS and the transitional state in between as the IS. 

An almost complete cycle in the CCD was traced during the outburst in 2004 May--June (Fig. 6, bottom-left). Before the flare the system was in the EIS. 
As the count rate in the HEXTE band started to decline, the PCA intensity went up triggering a fast jump of the system to the BS. Aql X-1 stayed in this 
X-ray state for about 10 days and then, again relatively suddenly,
within a couple of days, returned to the EIS via the IS. This kind of behaviour seems to be typical for many 
of the outbursts of Aql X-1 (see also \citealt{Rei04} who analyze in
some detail different outbursts than we do here). Although some
outbursts do not undergo a full cycle in the CCD, like the one on 2005 April (Fig. 6, upper-right), they do 
follow the same general trend in the CCD. The general pattern is the following: right before 
the onset of a major outburst in the PCA band the system is in the left
side of the EIS, migrates towards the right side of the EIS and as the
PCA intensity increases considerably it jumps suddenly to the BS. Here, moving towards its leftmost side (sometimes with excursions back to the right side), 
the peak of the PCA intensity is reached. Then, as the PCA intensity decreases the system jumps back to the EIS, passing quickly through the IS. 

\section{Correlations}

Although established only for a limited sample of objects, a correlation seems to exist, extending over many orders of magnitude, between the X-ray and
radio emissions of BHXRBs in the low-hard X-ray state \citep{Cor03,Cor08,Gal03,Gal06}. Recent evidence \citep{Xue07,Gal07} suggests that this correlation is not 
universal. More observations are definitely needed in order to settle the issue since given the relatively sparse nature of the data (particularly in 
the radio band) a confident 
conclusion cannot be reached at the moment. In the case of NSXRBs the situation is no better due to the fact that most of 
them systematically show weaker radio emission than BHXRBs (e.g. \citealt{Fen01b}). However, a correlation between the radio and X-ray emission has 
also been found for some of these objects in certain X-ray states, in particular the atoll source 4U 1728-34 \citep{Mig03,Mig06}.

\begin{center}
\begin{figure}
  \includegraphics[scale=0.33,angle=-90]{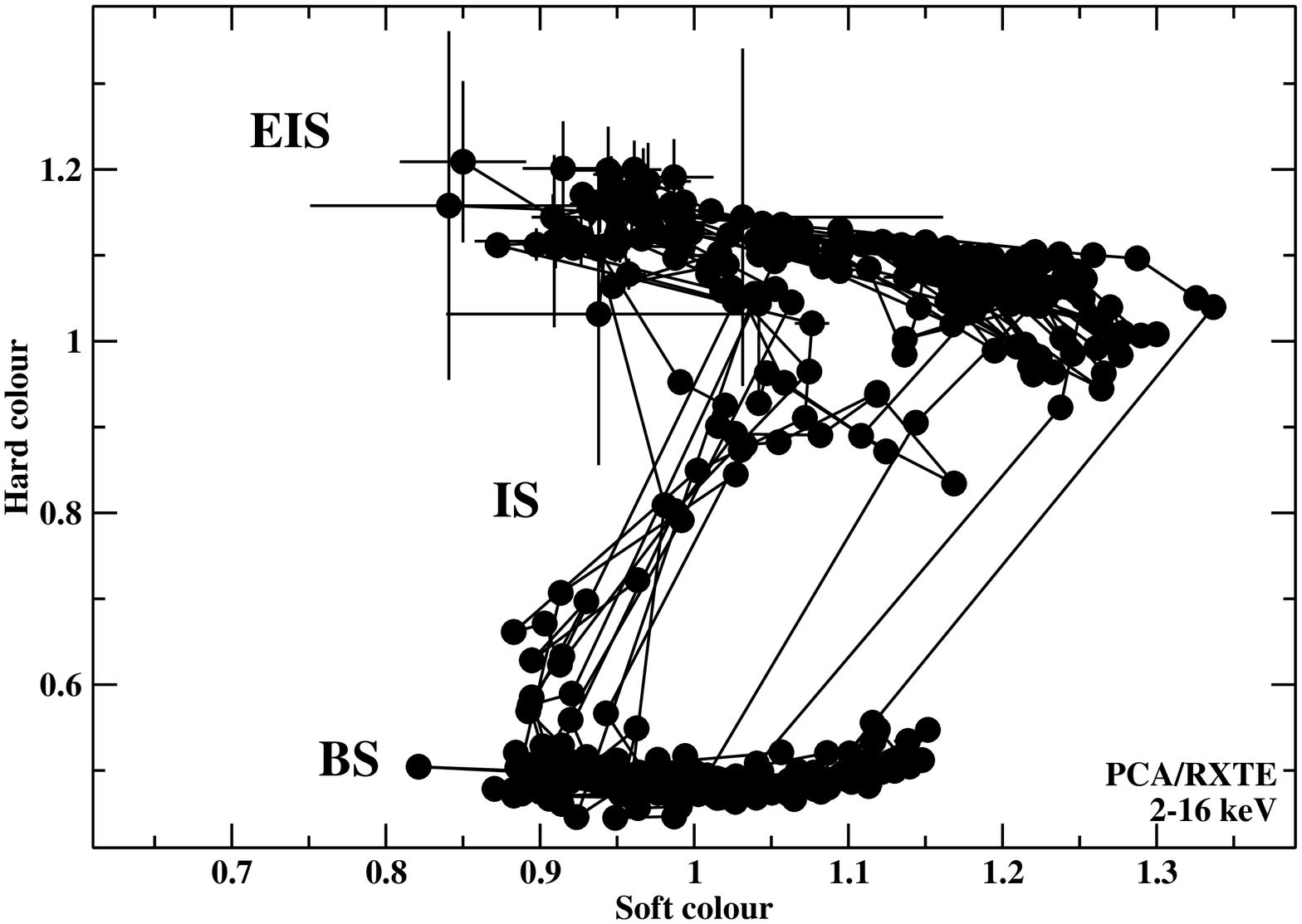}
  \caption{The cumulative CCD for all the outbursts of Aql X-1 observed with the PCA/RXTE between 1997--2008. While in the active state, the system 
spends most of the time in the ``extreme island state'' (EIS), with quick transitions through the ``island state'' (IS) and short excursions in the 
``banana state'' (BS). The soft colour is the ratio between the count rates in the 3.5--6.0 keV and 2.0--3.5 keV bands and the hard colour is the ratio 
between the count rates in the 9.7--16.0 keV and 6.0--9.7 keV bands.}
\end{figure}
\end{center}

For Aql X-1 we have found indications in the data that also for this object a correlation seems to exist between the radio and X-ray bands. Given the 
small number of observations under consideration this should be taken as tentative only, subject to confirmation or refutation when more data will be 
available (particularly in the radio band). Fig. 8 shows the radio flux density versus the PCA X-ray flux for Aql X-1 (data from Tables 1 and 2). 
When the system was in the EIS, the 8.4 GHz data (3 points only) reveal that the radio and X-ray bands have a positive correlation with a correlation 
coefficient of 0.78. The slope of the correlation function is
$\alpha$=0.29 $\pm$ 0.23 ($F_{radio} \propto F_{X}^{\alpha}$). Here and
further on, the errors on the correlation function are standard errors
of the regression fit. When considering 
only the 4.9 GHz data in the EIS (2 points only) the slope becomes
0.30. Taking into account all the radio detections in the EIS, independent of the frequency 
(i.e. 5 points), the data show no correlation between the radio and
X-ray bands (correlation coefficient 0.38). One possible explanation
for this, assuming the correlation at 8.4 GHz is real, is that the
spectrum between 4.9 and 8.4 GHz was not flat during these
observations. In BHXRBs for instance, 
in general in the low-hard state the radio spectrum is flat or optically thick (e.g. \citealt{Fen01a}), however in a few occasions it was observed that 
the spectrum changes rapidly and tends 
to become optically thin prior to a radio flare \citep{Fen04}. Perhaps something similar happens in Aql X-1. Unfortunately, in our data 
set the only constraints to the radio spectrum comes from the observations of the outbursts from 2002 March (Fig. 1) and 2004 May--June (Fig. 3), 
when the system was not in the EIS. When all the 8.4 GHz data are taken
into account, independent of the X-ray states of the system (i.e. 5 points), the slope
of the correlation function is $\alpha$=0.40 $\pm$ 0.07 with a
correlation coefficient of 0.96 (dashed line in Fig. 8). A similar analysis, for all the 4.9
GHz data (3 points) reveals an at best marginal correlation (correlation
coefficient 0.60) with a slope $\alpha$=0.48 $\pm$ 0.64.  

The radio upper limit denoted ``Q1'' in Fig. 8 is notable. It has a relatively high PCA flux and represents a deep radio upper limit. The 
system was in the BS at the time (point ``A'' in Fig. 4, top left). It looks as if the radio emission was ``quenched''. A very similar situation 
is observed in some BHXRBs in the high-soft state, where the radio emission drops significantly above some X-ray flux \citep{Gal03}. 

4U 1728-34 is the atoll source with the most numerous quasi-simultaneous detections in the radio and X-ray band (i.e. 10 up to now). We have used these 
data (VLA at 8.5 GHz with 100 MHz bandwidth, and PCA/RXTE in the
energy band 2--10 keV) reported in \cite{Mig03} and \cite{Mig06} to compare them against the Aql X-1 data from the point of view of the radio/X-ray correlations (Fig. 9). 
For Aql X-1, in calculating the luminosity we assumed the distance of 5.2
kpc used by \cite{Mig06}. A flat radio spectrum extending up to 8.5 GHz was also assumed. 
Following \cite{Mig03} and \cite{Mig06}, using all the data for 4U
1728-34 (except for the point denoted ``Q2'') we found a positive
correlation between the radio and X-ray bands, with the correlation coefficient 0.90 and the slope of the correlation 
function $\alpha$=1.40 $\pm$ 0.25 ($L_{radio} \propto L_{X}^{\alpha}$), in perfect agreement with previous authors. If moreover we consider only the data 
corresponding to the hard state (i.e. 7 
points; see \citealt{Mig06} for an identification of the X-ray states) the correlation coefficient becomes 0.96 and the slope 1.42 $\pm$ 0.20. It 
seems therefore that the correlation function for Aql X-1 is flatter than the corresponding correlation function for 4U 1728-34. 
The normalization of the trends however, appears to be similar. This 
result seems to be significant, although given the very sparse data
available at the moment in the radio band it is not possible to draw a
solid conclusion. More radio observations are stringently needed in order 
to further constrain the correlation between the radio and X-ray bands for these sources. 

As noted before by \cite{Mig03} the point ``Q2'' is interesting. It's the highest PCA X-ray flux at which the radio counterpart was detected, and the radio 
emission is relatively weak. Moreover, the system was perhaps in a soft
X-ray state, but this is not clear given the slightly contradictory
spectral and timing properties (see the discussion in \citealt{Mig03}). Thus, it might be that the radio emission was ``quenched'' during 
this observation. The radio detection ``Q2'' of 4U 1728-34 and the
radio upper limit ``Q1'' of Aql X-1 therefore suggest that NSXRBs might show a 
behaviour similar to BHXRBs with respect to the suppression of the radio emission above some X-ray flux levels. 

No significant correlation was found between the radio and optical bands. The correlation coefficient for the 3 observations in 2004 May--June for which 
quasi-simultaneous data exists at radio (8.4 GHz) and optical
wavelengths (see Table 1) in the EIS is 0.52. Adding also the observations in which the system 
was in the IS and the BS does not improve the correlation coefficient (0.01). 

\cite{Mai08} found correlations between the optical (R band) and ASM X-ray bands. The slope
of the correlation function seems to be steeper when the system is in the hard state, however outliers are present and it's not entirely clear what role, 
if any, the observational bias is playing. In turn, we were able to confirm this behaviour using PCA instead of ASM X-ray data 
(Fig. 10, top). All the multi-wavelength observations presented in Fig. 10 were quasi-simultaneous (separated by less than 1 day). The system in
the EIS follows a different optical/X-ray correlation than in the BS. 

Aql X-1 also shows distinct behaviours in the EIS and the BS X-ray states when the optical (R band) and HEXTE bands are compared (Fig. 10, middle). In the BS 
the HEXTE X-ray count rate is systematically low, independent of the
optical flux. This is contrary to the trend observed in the PCA band
where it is in the EIS that the system shows low X-ray intensities at very different optical flux levels (Fig. 10, top). 

The interplay between the X-ray emission in the PCA and HEXTE bands 
can be seen in Fig. 10, bottom. Please note that in relation to Fig. 10
the term ``correlation'' was used loosely, more like in the sense of
``trend''. Formally speaking, due to the 
scatter in the data, the correlation coefficients associated to Fig. 10 are not significant, with two exceptions: 0.93 for the optical/PCA X-ray 
bands in the BS (Fig. 10, top) and 0.95 for the HEXTE/PCA X-ray bands
in the EIS (Fig. 10, bottom). Since only data during outbursts are
available, and we only plotted the
quasi-simultaneous observations (thus only part of the total amount of data was selected), the analysis might be biased, in particular in 
the top and middle panels of Fig. 10. More quasi-simultaneous multi-wavelength observations are needed in order to constrain the behaviour of Aql X-1. 

\begin{center}
\begin{figure}
  \includegraphics[scale=0.33,angle=-90]{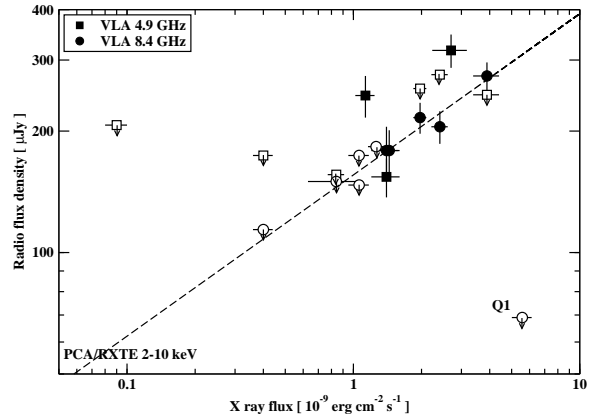}
  \caption{The radio/X-ray correlation in Aql
X-1. The filled circles and squares correspond to the radio detections
at 8.4 GHz and 4.9 GHz respectively. 
The empty symbols are the radio upper limits (3 $\sigma$). The dashed
line describes the correlation function (correlation coefficient 0.96)
when taking into account all the 8.4 GHz detections independent of the X-ray
state of the system, and has a slope $\alpha$=0.40 $\pm$ 0.07 ($F_{radio} \propto F_{X}^{\alpha}$). The 
radio upper limit denoted ``Q1'' is discussed in section 6.}
\end{figure}
\end{center}

\begin{center}
\begin{figure}
  \includegraphics[scale=0.33,angle=-90]{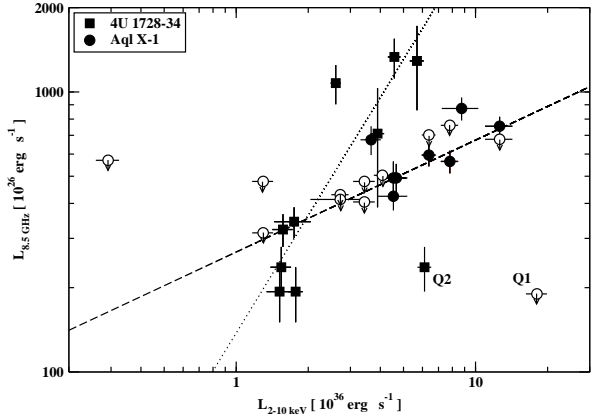}
  \caption{The radio (8.5 GHz)/X-ray (2--10 keV) correlations in Aql X-1 and 4U 1728-34. The filled circles and squares correspond to the radio detections of Aql X-1 and 
4U 1728-34 respectively. The empty circles are the radio upper limits
(3 $\sigma$) for Aql X-1. The dotted line is the correlation function (correlation coefficient 
0.90) for 4U 1728-34 (independent of the X-ray state) and has a slope
$\alpha$=1.40 $\pm$ 0.25 ($L_{radio} \propto L_{X}^{\alpha}$). The dashed line describes the 
correlation function (correlation coefficient 
0.96) for Aql X-1 when taking into account all the 8.4 GHz detections
independent of the X-ray state of the system and has a slope $\alpha$=0.40 $\pm$ 0.07. 
The data points denoted ``Q1'' and ``Q2'' are discussed in section 6.}
\end{figure}
\end{center}

\begin{center}
\begin{figure}
  \includegraphics[scale=0.33]{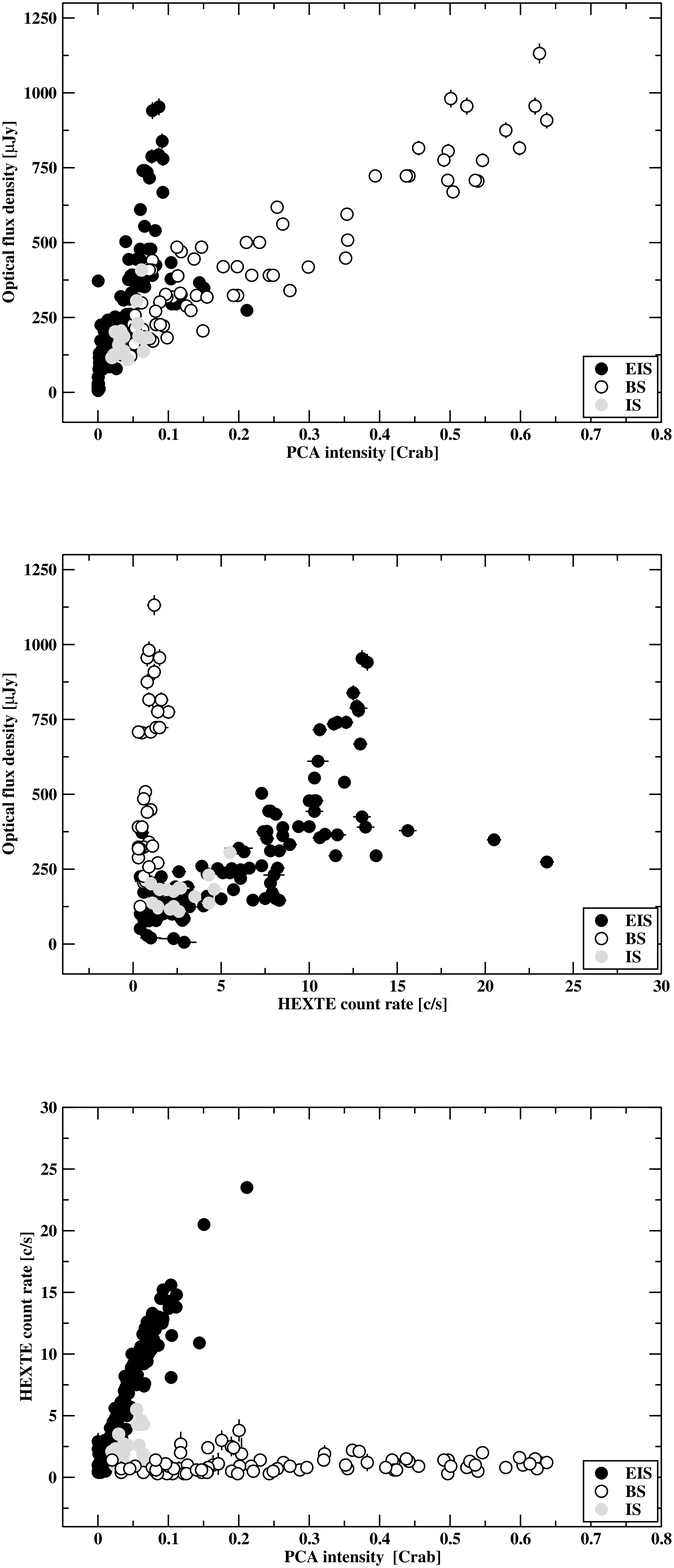}
  \caption{Trends in the behaviour of Aql X-1 in different X-ray
states. Filled black circles correspond to the ``extreme island state'' (EIS), open circles to 
the ``banana state'' (BS) and filled gray circles to the ``island state'' (IS). All the observations in different bands are quasi-simultaneous (separated by less 
than 1 day). The used energy bands are 2--16 keV for the PCA/RXTE and 20--200 keV for the HEXTE/RXTE. The optical band is the R band.}
\end{figure}
\end{center}

\section{Discussion and Conclusions}

Given the scarcity of the radio observations of Aql X-1 in the literature, we have searched the public VLA archive in the period 1986--2005 and found 
previously unreported detections thus bringing up the number of secure detections (more than 3 $\sigma$) from 5 \citep{Hje90,Rup04,Rup05} to 11. To 
attempt a multi-wavelength characterization of the system we used archival RXTE X-ray data (PCA and HEXTE) between 1997--2008 and the optical 
observations (R band) between 1998-2007 reported in \cite{Mai08}. 

In the combined data set on Aql X-1 we found 3 outbursts for which quasi-simultaneous radio, optical 
and X-ray data exist (Figs. 1, 2 and 3). These light curves look quite different from each other, at all the wavelengths, showing that the source 
is revealing a rich phenomenology. Judging by the hardness ratio, the three outbursts under discussion can be considered as 
corresponding to distinct classes of outbursts: in 2002 March the radio flare can be associated to a soft to hard transition; in 2004 May--June to a 
``classical'' hard to soft transition; and in 2005 April to a ``hard flare''. In the optical band \cite{Mai08} pointed out that the light curves of 
most of the outbursts of Aql X-1 do not belong to the pure fast rise exponential decay (FRED) topology and showed that FREDs occupy a particular 
region in the optical/X-ray plane that is distinct from the non-FRED outbursts.

But in spite of the diverse morphologies of the light curves, the outbursts of Aql X-1 tend to converge to a common behaviour when the HIDs and 
CCDs are investigated. The three outbursts presented in detail here, as well as most of the other events, track generally a similar path: 
counterclockwise in the HIDs and clockwise in the CCDs. There might be nonetheless exceptions. In one case during a minor outburst in the soft state 
the system 
tracked partially the HID (Fig. 5, the left side of the diagram) clockwise. During two or three minor outbursts the system was transiting 
between the hard and soft states and back again within a very small range in the X-ray intensity (Fig. 5, the bottom side of the diagram, between 
0.01--0.10 Crab) and 
it's hard to say with certainty whether it mimicked a scaled down version of the diagram or was doing something completely different. The 
fact remains that the HIDs of Aql X-1 exhibit similar characteristics to the HIDs of some BHXRBs (e.g. \citealt{Rod03,Ros04,Bel05,Dun08,Fen09}). For these 
objects a phenomenological model was developed that accounts for the connection between the radio and X-ray emission via the presence or 
absence of jets in the system \citep{Fen04,Fen09}. It goes as follows, starting from the lower right corner of the HID in Fig. 5. In the hard X-ray 
state a steady, optically thick jet is produced. During a typical outburst the X-ray intensity increases (moving up in the HID) and after a while the X-ray 
spectrum begins to soften (moving to the left in the HID). At the ``jet line''  the system enters the soft X-ray state when no jet is produced. The 
switching off of the jet is associated with relativistic 
ejections of matter, generally optically thin. The X-ray intensity decreases (moving down 
in the HID) and soon the X-ray spectrum starts to harden (moving right in the HID). At the crossing of the ``jet line'' the system switches on the 
steady jet and finishes the cycle back in the lower right corner of the HID. This kind of behaviour is roughly consistent 
with what is 
observed during the three outbursts of Aql X-1 (Fig. 4), but given the small number of radio detections it is practically impossible to test the 
model in more depth. The radio detection in a soft X-ray state with a likely inverted spectrum during the outburst from 2004 May--June 
(point ``H'' in Fig. 4, bottom left) does not necessarily contradict the idea of 
a turned off jet in the soft X-ray state since ad-hoc explanations are readily available, for instance by invoking a blob of matter ejected 
during the hard to soft transition at the ``jet line'', which is expanding adiabatically \citep{Laa66}.

The presence of hysteresis in Aql X-1 has already been established (e.g. \citealt{Mac03,Mai04,Gla07}) and confirmed in the present work (Fig. 5). 
Another atoll source, 4U 1608-522, shows a behaviour resembling that of
Aql X-1 (Linares \& van der Klis, in preparation). The similarity with the
hysteresis phenomenon observed in BHXRBs
(e.g. \citealt{Rod03,Ros04,Bel05,Dun08}) seems to suggest a common mechanism. 
What might be its nature is still a matter of debate. This is connected to the fact that at the moment, no self-consistent physical model of the X-ray 
states exists, although a few promising candidates are available (e.g. \citealt*{Rem06,Don07} for reviews). Therefore the trigger of the transitions between 
these states, which is likely the mechanism responsible for the hysteresis, is still an open issue (e.g. \citealt*{Don03,Zdz04,Mey05}).

Evidence for radio/X-ray correlations were found in Aql X-1 (Fig. 8). Taking into account the radio detections at 8.4 GHz, the slope 
of the correlation function is $\alpha$=0.40 $\pm$ 0.07 ($F_{radio} \propto F_{X}^{\alpha}$) with a correlation coefficient of 0.96. This is 
significantly different than the radio/X-ray correlation for another
atoll source, 4U 1728-34 \citep{Mig03,Mig06}, for which the data
indicate a slope of 1.40 $\pm$ 0.25 with the correlation coefficient
0.90, although the normalizations of the correlation functions 
are similar for both sources. However, these results are based on a very limited 
sample of data points, 5 in the case of Aql X-1 and 9 for 4U 1728-34, and therefore need further testing with future observations. Nevertheless, 
it's interesting to note that the slope of the correlations between the
radio (8.5 GHz) and X-ray bands (2--10 keV) observed in some BHXRBs in the hard state are
relatively close to what we obtained for Aql X-1: 0.51 $\pm$ 0.06 for V404 Cyg \citep{Cor08}, 0.58 $\pm$ 0.16 for 
A0620-00 \citep{Gal06}. But the normalizations of the correlation functions are higher for BHXRBs. This is related to the fact that the ratio between the radio 
and X-ray luminosity for BHXRBs is between one and two orders of magnitude higher than the corresponding ratio for NSXRBs \citep{Fen01b,Mig06}. This can 
be explained for instance by a more efficient (i.e. ``radio loud'') jet
production mechanism in BHXRBs than in NSXRBs \citep{Fen03} which might
come about due to the
fact that the geometrically thick disks are more efficient at producing
outflows (e.g. \citealt*{Mei01}) and the disks of NSXRBs are truncated at
larger radii than those of BHXRBs. Or it 
might be that 
in the case of BHXRBs a large fraction of the extracted gravitational energy is advected across the event horizon, the end product of the process being more 
``X-ray quiet'' systems (e.g. \citealt{Nar97}). From a more general point of view, it was 
shown that a scaling relation between the X-ray (2--10 keV) and radio
(5 GHz) bands with a slope of 0.60 $\pm$ 0.11 holds for a fairly large sample of BHXRBs and 
active galactic nuclei (AGN) when the mass of the compact objects is
taken into account \citep*{Mer03,Fal04}. Given also the recent
advancements (e.g. the relation between the mass of the black hole, the
accretion rate and the break frequency in X-ray power spectra for AGNs
and XRBs [\citealt{Har06}], or the similarity between the HIDs of XRBs and the HID
of a dwarf nova in outburst [\citealt{Kor08}]), it seems that evidence is
mounting at a fast pace that the accretion/ejection process works in a
similar way in a broad class 
of objects: stellar mass black holes in XRBs, super-massive black holes in AGNs, and perhaps neutron stars in XRBs and white dwarfs in cataclysmic variables. 

Aql X-1 shows evidence for the ``quenching'' of the radio emission in the soft X-ray state, above some X-ray flux (the radio upper limit ``Q1'' 
in Fig. 9). \cite{Mig03} found a similar behaviour in 4U 1728-34 (``Q2'' in Fig. 9). Although it's premature to jump to conclusions with such a 
limited sample, these are at least good indications that NSXRBs could mimic the behaviour of some BHXRBs from this point of view.

\cite{Mai08} found a correlated behaviour between the optical (R band) and X-ray bands depending on the state of the system. Using the same optical 
data set, and PCA X-ray data instead of ASM, we confirmed their findings. Moreover, we showed that trends exist also between the optical (R band) 
and HEXTE X-rays bands, and between the PCA and HEXTE bands (Fig. 10). \cite{Mai08} proved that during the outbursts the evolution of the 
optical/infrared colour magnitude diagrams of Aql X-1 is consistent with thermal heating of an irradiated outer accretion disc. They noted that the 
optical/infrared 
colour and flux do not seem to change suddenly during the X-ray state transitions. This can be interpreted as evidence against a non-thermal 
origin of the optical/infrared emission. We didn't find any significant correlation between the optical (R band) and radio emission 
(at cm wavelengths), and although based on a very limited number of
radio observations, this adds further support to this interpretation. 

As a general remark, our data are roughly consistent with a truncated disc model of the X-ray states (e.g. \citealt{McC06} for a review). In hard 
X-ray states a standard accretion disc is truncated relatively far from the compact object. In soft X-ray states the 
disc is advancing closer to the innermost stable circular orbit. In the soft states the radiation is well described by blackbody thermal emission 
from the inner regions of the accretion disc. But in the hard states the nature of the emitting material filling the gap between the compact object and 
the truncated disc (generically called the ``corona'') is still under debate (e.g. \citealt{Esi01,Mar01,Bla04}), although it is quite 
clear that synchrotron and Compton mechanisms are important. The fact that the optical emission (R band) does not seem to change significantly in 
the two X-ray states is further circumstantial evidence for the interpretation of \cite{Mai08} of its origin as irradiation of the outer accretion disc. 

\section*{Acknowledgments}

We thank the anonymous referee for detailed comments that improved the
manuscript. The National Radio Astronomy Observatory is a facility of the National Science Foundation operated under cooperative 
agreement by Associated Universities, Inc.. This research has made use
of data obtained through the High Energy Astrophysics Science Archive
Research Center Online Service, provided by the NASA Goddard Space
Flight Center.

\end{document}